\documentstyle[epsfig,twocolumn]{mn}

\title[A time\,-\,luminosity correlation for GRBs in the X\,-\,rays]
{A time\,-\,luminosity correlation for Gamma - Ray Bursts in the X\,-\,rays}
\author[M.G. Dainotti et al.]{M.G. Dainotti$^{1,2}$, V.F. Cardone$^{3,4}$, S. Capozziello$^{4,5}$\\
$^1$ICRANet and ICRA, Piazzale della Repubblica 10, 65122 Pescara,
Italy \\ $^2$ Dipartimento di Fisica, Universit\`{a} di Roma "La
Sapienza", Piazzale Aldo Moro 5, 00185 Roma, Italy \\
$^3$ I.N.A.F. - Osservatorio Astrofisico di Catania, via Santa
Sofia 78, 95123 - Catania, Italy \\ $^4$ Dipartimento di Scienze
Fisiche, Universit\`{a} di Napoli "Federico II", Complesso
Universitario di Monte Sant' Angelo, \\ Edificio N, via Cinthia,
80126 - Napoli, Italy
\\ $^5$ I.N.F.N., Sez. di Napoli, Complesso Universitario di Monte
Sant' Angelo, Edificio G, via Cinthia, 80126 - Napoli, Italy}

\date{Accepted xxx, Received yyy, in original form zzz}

\begin{document}
\maketitle

\begin{abstract}
Gamma ray bursts (GRBs) have recently attracted much attention as a
possible way to extend the Hubble diagram to very high redshift. However,
the large scatter in their intrinsic properties prevents directly using
them as distance indicator so that the hunt is open for a relation
involving an observable property to standardize GRBs in the same way as the
Phillips law makes it possible to use Type Ia Supernovae (SNeIa) as
standardizable candles. We use here the data on the X\,-\,ray decay curve
and spectral index of a sample of GRBs observed with the Swift satellite.
These data are used as input to a Bayesian statistical analysis looking for
a correlation between the X\,-\,ray luminosity $L_X(T_a)$ and the time
constant $T_a$ of the afterglow curve. We find a linear relation between
$\log{[L_X(T_a)]}$ and $\log{[T_a/(1+z)]}$ with an intrinsic scatter
$\sigma_{int} = 0.33$ comparable to previously reported relations.
Remarkably, both the slope and the intrinsic scatter are almost independent
on the matter density $\Omega_M$ and the constant equation of state $w$ of
the dark energy component thus suggesting that the circularity problem is
alleviated for the $L_X - T_a$ relation.

\end{abstract}


\section{Introduction}

The high fluence values (from $10^{-7}$ to $10^{-5} \ {\rm  erg/cm^2}$) and
the enormous isotropic energy emitted ($\simeq 10^{50} - 10^{54} {\rm
erg}$) at the peak in a single short pulse make Gamma Ray Bursts (hereafter
GRBs) the most violent and energetic astrophysical phenomena.
Notwithstanding the variety of their different peculiarities, some common
features may be identified looking at their light curves. Although GRBs
have been traditionally classified as {\it short} and {\it long} depending
on $T_{90}$ being smaller or larger than $2 \ {\rm s}$ (with $T_{90}$ the
time over which from $5\%$ to $95\%$ of the prompt emission is released), a
recent analysis by Donaghy et al. \shortcite{donaghy} has shown that this
criterion has to be revised. Indeed, the existence of an intermediated
class of GRBs have also been studied \cite{nb06,Bernardini07}. As a result,
the long GRBs are now further classified as {\it normal} and {\it low
luminosity} with the latter ones probably associated with Supernovae
\cite{pian06,Dainotti}.

Notwitstanding this classification, two phases are clearly visible in the
GRB lightcurve, namely the prompt emission, where most of the energy is
released in the $\gamma$\,-\,rays in only tens of seconds, and an afterglow
lasting many hours after the initial bursts. Early observations in the
X\,-\,rays typically started several hours after the prompt emission so
that only the late phase of the light curve could be characterized. It was
then found that a phenomenological power\,-\,law, $f(t, \nu) \propto
t^{-\alpha} \nu^{-\beta}$ with $(\alpha, \beta) \simeq (-1.4, 0.9)$,
provided a reasonable fit to the observed data \cite{Piro01}. However, the
launch of the {\it Swift} satellite, whose aim is also to observe GRBs
X\,-\,ray ($0.2 - 10 {\rm keV}$) and optical ($1700 - 6500$ {\AA}) afterglows
starting few seconds after the trigger, revealed a more complex behaviour.
The soft X\,-\,ray light curves must indeed be divided in two different
classes \cite{Chinca05} according to the steep or mild initial decay. Most
of the observed GRB afterglows belong to the first group, showing what has
been called a {\it canonical} behavior \cite{nousek06} described by a
broken power\,-\,law. After the initial steep decay (with slope $3 \leq
\alpha_1 \leq 5$), the light curve shows a shallow decay ($0.5\leq
\alpha_2 \leq 1$) followed by a somewhat steeper decay ($1 \leq
\alpha_3 \leq 1.5$) beyond $2 {\times} 10^4 \ {\rm s}$. These
power\,-\,law segments are separated by two corresponding break times with
$t_{b1} \leq 500 \ {\rm s}$ and $10^{3} \ {\rm s} \leq t_{b2} \leq 10^{4} \
{\rm s}$. A new systematic study using GRBs observed with XRT reveals a
still more complex behavior with different power\,-\,law slopes and break
times \cite{OB06,Sak07}. A significant step forward has been represented by
the analysis of the X\,-\,ray afterglow curves of the full sample of {\it
Swift} GRBs showing that all of them may be fitted by the same analytical
expression \cite{W07}.

Finding out a universal feature for GRBs is the first important step
towards their use as distance indicator. To this aim, one has indeed to
look for a universal relation linking observable GRBs properties so that
their intrinsic luminosity may be estimated from directly measurable
quantities. Previous attempts along this road are represented by the $E_iso
- E{peak}$ \cite{Amati06}, $E_\gamma - E_{peak}$ \cite{G04,Ghirlanda06},
$L -E_{peak}$ \cite{S03}, $L - \tau_{lag}$ \cite{N00}, $L - V$
\cite{FRR00,R01}, $L - \tau_{RT}$. Moreover, three\,-\,parameter relations
have also been proposed such as, e.g., the $E_iso - E_p - t_b$
\cite{liza05} and that proposed by Firmani and collaborators (Firmani et
al. 2005, 2006). On the other hand, some attempts have also been made to
compare these empirical correlations with the model dependent ones
\cite{na06,Guida08}. The above quoted two parameters correlations have then
been used by Schaefer (2007, hereafter S07) to construct the first reliable
GRBs Hubble diagram extending up to $z \simeq 6$ opening the way towards
the use of GRBs as cosmological probes (see, e.g., Capozziello \& Izzo 2008
and refs. therein).

In this letter, we present a possible alternative route towards
standardizing GRBs as distance indicator. To this aim, we use the data in
Willingale et al. (2007) to look for a possible correlation between the
X\,-\,ray luminosity at the break time $T_a$ and the $T_a$ itself. The data
used are presented in Sect. 2, while Sect. 3 deals with the statistical
tools and the results. Conclusions are summarized in Sect. 4.

\section{The data}

Willingale et al. (2007, hereafter W07) have examined the
X\,-\,ray decay curves of all the GRBs measured by the {\it Swift}
satellite then available. Their analysis shows that all of them
may be well fitted by a simple two components formula, namely\,:

\begin{equation}
f(t) = f_p(t) + f_a(t)
\label{eq: fluxtot}
\end{equation}
where the first term accounts for the prompt $\gamma$\,-\,ray emission and
the initial X\,-\,ray decay, while the second one describes the afterglow.
Both components are given by the same functional expression\,:

\begin{equation}
f_c(t) = \left \{
\begin{array}{ll}
\displaystyle{F_c \exp{\left ( \alpha_c - \frac{t \alpha_c}{T_c} \right )} \exp{\left (
- \frac{t_c}{t} \right )}} & {\rm for} \ \ t < T_c \\
~ & ~ \\
\displaystyle{F_c \left ( \frac{t}{T_c} \right )^{-\alpha_c}
\exp{\left ( - \frac{t_c}{t} \right )}} & {\rm for} \ \ t \ge T_c \\
\end{array}
\right .
\label{eq: fc}
\end{equation}
where the transition from the exponential to the power,\-\,law
decay takes place at the point $(T_c, F_c)$ where the two
functional sections match in value and gradient. The parameter
$\alpha_c$ determines both the time constant of the exponential
decay (given by $T_c/\alpha_c$) and the slope of the following
decay, while $t_c$ marks the initial rise and the time of maximum
flux occuring at $t = \sqrt{t_c T_c/\alpha_c}$. Denoting with the
suffix $p$ and $a$ quantities for the prompt and afterglow
components, Eq.(\ref{eq: fc}) may be inserted into Eq.(\ref{eq:
fluxtot}) to give an eight parameters expression that can be
fitted to the X\,-\,ray decay curve in order to both validate this
expression and determine, for each GRB, the corresponding
parameters. Such a task has been indeed performed by W07 using all
the 107 GRBs detected by both BAT and the XRT on {\it Swift} up to
August 1st 2006. The fit procedure and the detailed analysis of
the results are presented in W07, while here we only remind that
the usual $\chi^2$ fitting in the $\log{(flux)}$ vs $\log{(time)}$
provide estimates and uncertainties on the time parameters
$(\log{T_p}, \log{T_a})$ and the products $(\log{F_p T_p},
\log{F_a T_a})$.

W07 also performed spectral fitting with XSPEC \cite{xspec} to BAT
(for the prompt phase) and XRT (for later phases) data to estimate
the spectral index during different phases. Due to the limited
frequency range, the GRB spectrum may be simply described by a
single power\,-\,law, $\Phi(E) \propto E^\beta$, with the slope
$\beta$ depending on the time when the spectrum is observed. W07
reported four different values of $\beta$, namely $\beta_p$ (for
the prompt phase), $\beta_{pd}$ for the prompt decay, $\beta_a$
for the plateau observed at the time $T_a$, and $\beta_{ad}$ for
the afterglow at $t > T_a$. Actually, the data coverage is not
sufficient to measure all of them for the full sample so that, for
the weakest bursts, only $\beta_p$ and $\beta_{pd}$ are available.
Provided $\beta$ is known, it is possible to estimate the GRB
luminosity at a given time $t$ as\,:

\begin{equation}
L_X(t) = 4 \pi D_L^2(z) F_X(t)
\label{eq: lx}
\end{equation}
where $D_L(z)$ is the luminosity distance at the GRB redshift $z$,
and $F_X(t)$ is the flux (in ${\rm erg/cm^2/s}$) at the time $t$,
$K$\,-\,corrected \cite{B01} as\,:

\begin{equation}
F_X(t) = f(t) \ {\times} \
\frac{\int_{E_{min}/(1 + z)}^{E_{max}/(1 + z)}{E \Phi(E)
dE}}{\int_{E_{min}}^{E_{max}}{E \Phi(E) dE}}
\label{eq: fx}
\end{equation}
with $(E_{min}, E_{max}) = (0.3, 10) \ {\rm keV}$ set by the
instrument bandpass. Note that Eq.(\ref{eq: lx}) is the same as
Eq.(8) in S07 the only difference being the integration limits of
the integral at the numerator. Actually, while S07 is interested
to the bolometric luminosity, we are here concerned with the
X\,-\,ray one so that we integrate only over this energy range.

Using the data in W07, we compute the X\,-\,ray luminosity at the
time $T_a$ so that we have to set $f(t) = f(T_a)$ and $\beta =
\beta_a$ in Eqs.(\ref{eq: lx}) and (\ref{eq: fx}). Actually,
rather than using Eq.(\ref{eq: fluxtot}), we set $f(T_a) =
f_a(T_a)$ since the contribution of the prompt component is
typically smaller than $5\%$, much lower than the statistical
uncertainty on $f_a(T_a)$. Neglecting $f_p(T_a)$ thus allows to
reduce the error on $F_X(T_a)$ without introducing any bias. This
latter error is then estimated by simply propagating those on
$\beta_a$, $\log{T_a}$ and $\log{F_a T_a}$ thus implicitly
assuming that their covariance is null\footnote{Note that the
covariance matrix is not reported in W07, where the parameters of
interest are given with their $90\%$ confidence ranges. Following
Willingale (priv. comm.), we have assumed independent Gaussian
errors and obtained $1 \sigma$ uncertainties by roughly dividing
by 1.65 the $90\%$ errors. Moreover, we preliminary correct for
asymmetric errors on $\log{F_a T_a}$ and $\log{T_a}$ (when
present) following the prescriptions in D' Agostini (2004).}.
Should this not be the case, we are underestimating the final
error on $L_X(T_a)$. We have, however, checked that our main
results are unaffected by a reasonable increase of the errors.

As a final important remark, we note that the presence of the
luminosity distance $D_L(z) = (c/H_0) d_L(z)$ in Eq.(\ref{eq: lx})
constrains us to adopt a cosmological model to compute $L_X(T_a)$.
We use a flat $\Lambda$CDM model so that the Hubble free
luminosity distance reads\,:

\begin{equation}
d_L(z) = (1 + z) \int_{0}^{z}{\frac{dz'}{\sqrt{\Omega_M (1 + z')^3 +
(1 - \Omega_M)}}} \ .
\label{eq: dl}
\end{equation}
In agreement with the WMAP five year results \cite{WMAP5}, we set
$(\Omega_M, h) = (0.291, 0.697)$ with $h$ the Hubble constant
$H_0$ in units of $100 \ {\rm km/s/Mpc}$.

\section{A luminosity\,-\,time correlation}

In order to standardize GRBs to use them as possible distance
indicator, we need to find a correlation between the luminosity
and a directly observable quantity. Should such a relation be
found, one can then use the observed flux and the estimated $L_X$
to infer $D_L(z)$ and then construct the GRB Hubble diagram. Let
us suppose that a power\,-\,law relation exists between two
quantities $R$ and $Q$ as $R = A Q^B$. In logarithmic units, this
reads $ \log{R} = a + b \log{Q}$ with $a = \log{A}$ and $b = B$.
Typically, both $R$ and $Q$ will be known with measurement errors
($\sigma_R, \sigma_Q)$ so that the statistical uncertainties on
$(\log{R}, \log{Q})$ will be given by $(\sigma_R/R, \sigma_Q/Q)
{\times} (1/\ln{10})$ respectively. These errors may be comparable
so that it is not possible to decide what is the independent
variable to be used in the usual $\chi^2$ fitting analysis.
Moreover, the relation $R = A Q^B$ may be affected by an intrinsic
scatter $\sigma_{int}$ of unknown nature that has to be taken into
account. In order to determine the parameters $(a, b,
\sigma_{int})$, we can then follow a Bayesian approach
\cite{Dago05} thus maximizing the likelihood function
${\cal{L}}(a, b, \sigma_{int}) = \exp{[-L(a, b, \sigma_{int})]}$
with\,:

\begin{eqnarray}
L(a, b, \sigma_{int}) & = &
\frac{1}{2} \sum{\ln{(\sigma_{int}^2 + \sigma_{y_i}^2 + b^2
\sigma_{x_i}^2)}} \nonumber \\
~ & + & \frac{1}{2} \sum{\frac{(y_i - a - b x_i)^2}{\sigma_{int}^2 + \sigma_{y_i}^2 + b^2
\sigma_{x_i}^2}}
\label{eq: deflike}
\end{eqnarray}
with $(x_i, y_i) = (\log{Q_i}, \log{R_i})$ and the sum is over the
${\cal{N}}$ objects in the sample. Note that, actually, this maximization
is performed in the two parameter space $(b, \sigma_{int})$ since $a$ may
be estimated analytically as\,:

\begin{equation}
a = \left [ \sum{\frac{y_i - b x_i}{\sigma_{int}^2 + \sigma_{y_i}^2 + b^2
\sigma_{x_i}^2}} \right ] \left [\sum{\frac{1}{\sigma_{int}^2 + \sigma_{y_i}^2 + b^2
\sigma_{x_i}^2}} \right ]^{-1}
\label{eq: calca}
\end{equation}
so that we will not consider it anymore as a fit parameter.

\begin{figure}
\includegraphics[width=8.5cm]{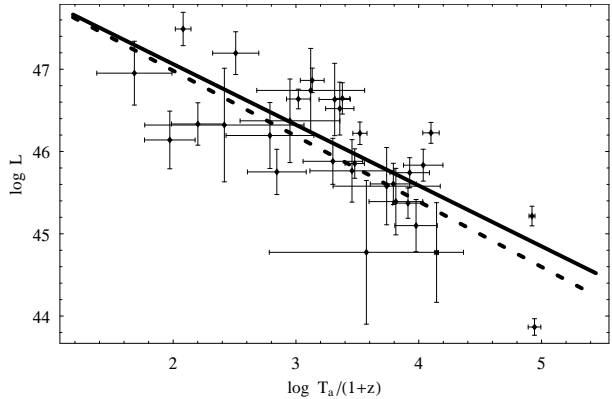}
\label{fig: bfallplot} \caption{Best fit curves superimposed to
the data with the solid and dashed lines referring to the results
obtained with the Bayesian and Levemberg\,-\,Marquardt estimator
respectively.}
\end{figure}

We use this general recipe to look for a correlation between the X\,-\,ray
luminosity (in ${\rm erg \ s^{-1}}$) at the time $T_a$ and $T_a$ (in ${\rm
s}$) itself, i.e. we set $y = \log{[L_X(T_a)]}$ and $x =
\log{[T_a/(1+z)]}$, where we divide time by $(1 + z)$ to account for the
cosmological time dilation. Note that, since $\beta_a$ is needed to compute
$L_X(T_a)$, we have to reject most of the 107 GRBs reported in W07 because
this is not known. We thus end up with a sample contanining ${\cal{N}} =
32$ with both $\log{[L_X(T_a)]}$ and $\log{[T_a/(1+z)]}$
measured\footnote{ASCII tables with all the quantities needed for the
analysis and the {\it Mathematica} codes used are available on request.}.
The Spearman rank correlation turns out to be $r = -0.74$ suggesting that a
power\,-\,law relation between $L_X(T_a)$ and $T_a/(1+z)$ indeed exists
thus motivating further analysis.

We then apply the maximum likelihood estimator described above in order to
determine both the slope and the intrinsic scatter of the $L_X - T_a$
correlation thus finding out\,:

\begin{displaymath}
(a, b, \sigma_{int}) = (48.54, -0.74, 0.43) \ .
\end{displaymath}
Defining the best fit residuals as $\delta = y_{obs} - y_{fit}$, we can
qualitatively estimate the goodness of the fit by considering the median
and root mean square which turn out to be $\langle \delta \rangle = -0.08$
and $\delta_{rms} = 0.52$ indeed quite small if compared to the typical
$\log{[L_X(T_a)]}$ values. It is also worth noting that $\delta$ does not
correlate with the other parameters of the fit flux, while the value $r =
-0.23$ between $\delta$ and $z$ favours no significative evolution of the
$L_X - T_a$ relation with the redshift. The best fit relation is
superimposed to the data in Fig.\,1 where we also present the best fit
obtained by the usual $\chi^2$ fitting technique. In this case, the best
fit parameters are obtained by minimizing (through a
Levemberg\,-\,Marquardt algorithm with $1.5 \sigma$ outliers rejection) a
$\chi^2$ merit function given by the second term in Eq.(\ref{eq: deflike})
with $\sigma_{y_i} = \sigma_{int} = 0$, i.e. we (erroneously) assume that
there is no scatter and that the errors on $\log{[L_X(T_a)]}$ are
negligible. This alternative method gives as best fit parameters\,:

\begin{displaymath}
(a, b) = (48.58, -0.79)
\end{displaymath}
in good agreement with the above maximum likelihood estimator so that we
argue that our results are independent on the fitting method. However,
since the Bayesian approach is better motivated and also allows for an
intrinsic scatter, we hereafter elige this as our preferred technique.

\begin{figure}
\includegraphics[width=8.5cm]{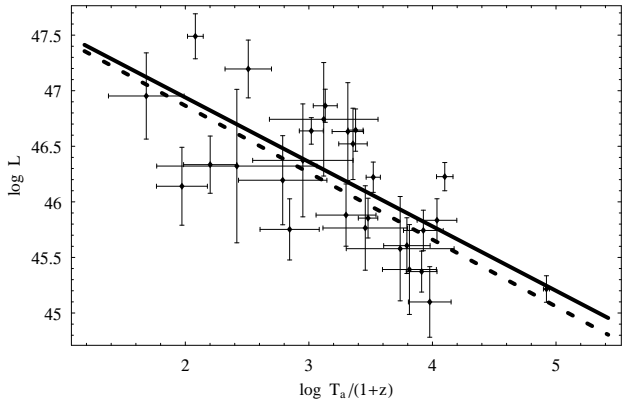}
\label{fig: bfbestplot} \caption{Same as Fig.\,\ref{fig:
bfallplot}, but only using GRBs with $1 \le \log{[T_a/(1+z)]} \le 5$ and
$\log{[L_X(T_a)]} \ge 45$.}
\end{figure}

In an attempt to reduce the intrinsic scatter in the above correlation, we
have analysed the best fit residuals noting that the higher ones are
obtained for GRBs with luminosities smaller than $10^{45} \ {\rm erg}$ and
time parameter $\log{[T_a/(1+z)]}> 5$. We therefore repeated the above
analysis using only 28 out of 32 GRBs\footnote{The four GRBs excluded
are\,: GRB050824, GRB060115, GRB060607A and GRB060614. While the first two
appear to be unaffected by any problem, for the latter two, the data cover
less than $50\%$ of the $T_{90}$ range . Moreover, for GRB060607A, the
prompt component dominates over the afterglow one so that our approximation
$f(T_a) \simeq f_a(T_a)$ is not valid anymore.} satisfying the two
selection criteria $1 \le \log{[T_a/(1+z)]} \le 5$  and ${[L_X(T_a)]} \ge
45$. Using the maximum likelihood estimator, we get\,:

\begin{displaymath}
(a, b, \sigma_{int}) = (48.09, -0.58, 0.33)
\end{displaymath}
with $\langle \delta \rangle = -0.06$ and $\delta_{rms} = 0.43$. The
reduced intrinsic scatter and the smaller fit residuals suggest us that,
whatever is the unknown mechanism originating the $L_X - T_a$ relation,
this is better effective for the class of GRBs satisfying the above
selection criteria. The data and the best fit curve are shown in Fig.\,2
where the dashed line refers to the results obtained with the $\chi^2$
minimization giving $(a, b) = (48.07, -0.60)$ reported here for
completeness.

The Bayesian approach used here also allows us to quantify the
uncertainties on the fit parameters. To this aim, for a given
parameter $p_i$, we first compute the marginalized likelihood
${\cal{L}}_i(p_i)$ by integrating over the other parameter. The
median value for the parameter $p_i$ is then found by solving\,:

\begin{equation}
\int_{p_{i,min}}^{p_{i,med}}{{\cal{L}}_i(p_i) dp_i} = \frac{1}{2}
\int_{p_{i,min}}^{p_{i,max}}{{\cal{L}}_i(p_i) dp_i} \ .
\label{eq: defmaxlike}
\end{equation}
The $68\%$ ($95\%$) confidence range $(p_{i,l}, p_{i,h})$ are then found by
solving\,:

\begin{equation}
\int_{p_{i,l}}^{p_{i,med}}{{\cal{L}}_i(p_i) dp_i} = \frac{1 - \varepsilon}{2}
\int_{p_{i,min}}^{p_{i,max}}{{\cal{L}}_i(p_i) dp_i} \ ,
\label{eq: defpil}
\end{equation}

\begin{equation}
\int_{p_{i,med}}^{p_{i,h}}{{\cal{L}}_i(p_i) dp_i} = \frac{1 - \varepsilon}{2}
\int_{p_{i,min}}^{p_{i,max}}{{\cal{L}}_i(p_i) dp_i} \ ,
\label{eq: defpih}
\end{equation}
with $\varepsilon = 0.68$ (0.95) for the $68\%$ ($95\%$) range
respectively. For the fit to the full dataset, we get\,:

\begin{displaymath}
b = -0.74_{-0.19 \ -0.39}^{+0.20 \ +0.41} \ \ , \ \
\sigma_{int} = 0.48_{-0.10 \ -0.18}^{+0.15 \ +0.35} \ \ ,
\end{displaymath}
while it is\,:

\begin{displaymath}
b = -0.58_{-0.18 \ -0.37}^{+0.18 \ +0.38} \ \ , \ \
\sigma_{int} = 0.39_{-0.11 \ -0.20}^{+0.14 \ +0.33} \ \
\end{displaymath}
for the selected subsample.

\section{Discussion and conclusion}

The high Spearman correlation coefficient, the low value of the fit
residuals and the modest intrinsic scatter renders the $L_X - T_a$ relation
presented above a new valid tool to standardize GRBs. It is worth stressing
that $L_X - T_a$ needs only two parameters and one of them is directly
inferred form the observations minimizing the effects of the systematics
errors. Furthermore the redshift range covered is large extending from
$0.54$ $(0.125)$ up to 6.6 for the selected (full) sample far beyond the
maximum redshift affordable with Type Ia SNe ($z \approx 1.7$). Should this
correlation be confirmed by future higher quality data, one could then
combine it with the other relations yet available in literature to work out
a GRBs Hubble diagram deep into the matter dominated era thus representing
an outstanding cosmological test.

To this end, it is worth comparing the $L_X - T_a$ relation with other ones
quoted in literature. When performing such a comparison, however, one
should take into account the differences in the cosmological model adopted
and the fitting method used. In particular, the choice of how the best fit
parameters are estimated may have an important impact on the estimate of
the intrinsic scatter with the usual $\chi^2$ fitting leading to an
underestimate of $\sigma_{int}$. On the other hand, changing $\Omega_M$ in
the framework of the flat $\Lambda$CDM scenario have a profound impact on
$\sigma_{int}$ with higher $\Omega_M$ giving rise to lower $\sigma_{int}$
values \cite{BP08}. In order to account for both these issues, one should
therefore test all the above correlations using the same statistical tools
and cosmological model, a task we will address elsewhere.

As is well known, the paucity of local (i.e., $z \le 0.1$) GRBs represents
a serious problem for any attempt to standardize GRBs since it is very
difficult to directly calibrate any relation. This problem may be partly
overcome by fitting the correlation in a subsample of GRBs lying at similar
redshift \cite{Amati08}. However, as a general rule, in order to evaluate
the GRB luminosity, a cosmological model has to be adopted thus leading to
the circularity problem. Although addressing this problem in detail will be
the subject of a forthcoming work, we have here investigated what is the
effect of changing the cosmological model by using our maximum likelihood
estimator to determine the parameters $(a, b, \sigma_{int})$ as function of
$\Omega_M$ in a flat $\Lambda$CDM model\footnote{To be precise, we let
$\Omega_M$ running from $0.2$ to $1$ and adjust $h$ so that $\Omega_M h^2$
is fixed to the same value adopted above.}. We find the remarkable result
that both the best fit parameters $(b, \sigma_{int})$ and the rms residual
$\delta_{rms}$ are almost insensitive to the value of $\Omega_M$. Indeed,
$b$ runs from $b \simeq -0.590$ to $b \simeq -0.565$, while $\sigma_{int}$
increases from $\sigma_{int} \simeq 0.335$ to $\sigma_{int} \simeq 0.340$
for $\Omega_M$ going from 0.2 to 1.0. As a further test, we generalize the
$\Lambda$CDM model varying not only the matter density parameter
$\Omega_M$, but also the equation of state $w$ of the dark energy component
(with $w = -1$ for the $\Lambda$CDM model). For $-1.3 \le w \le -0.7$,
neither $b$ nor $\sigma_{int}$ significantly change confirming the
qualitative results obtained for the $\Lambda$CDM scenario. Although a more
detailed analysis is needed, we therefore argue that the circularity
problem is alleviated by the use of our $L_X - T_a$ relation.

The encouraging results discussed above are serious arguments in favour of
the $L_X - T_a$ relation as a further tool towards the standardization of
GRBs as distance indicator. Should these first evidences be furtherly
enforced by more data, the combined use of full set of GRBs correlations
discovered insofar could opened the road towards making GRBs the high
redshift analog of SNeIa as cosmological probes in the not too distant
future. \\

{\it Acknowledgements.} We warmly thank R. Willingale for help with the
data and prompt answers to our questions and an anonymous referee for
his/her valuable comments.

\end{document}